\documentclass[letter]{aa}     

\usepackage{graphicx}
\usepackage{txfonts}
\usepackage{lipsum}
\usepackage{subcaption}         
\usepackage{lscape}             
\usepackage{placeins}           
\usepackage{natbib}                               
\usepackage[colorlinks=true,linkcolor=bluea,allcolors=blue]{hyperref}

\newcommand{\kms}{km\,s$^{-1}$}
\begin{document}

 \title{Extended coronal line emission and new clues to a possible dual AGN in the merger J1356+1026}


   \author{M. Bianchin\inst{1,2}\thanks{\email{marina.bianchin@iac.es}}
        \and C. Ramos Almeida\inst{1,2}
        \and O. González-Martín\inst{3}
        \and M. V. Zanchettin\inst{4}
        \and M. Carneiro\inst{5}
        \and M. Pereira-Santaella\inst{6}
        \and C. Tadhunter\inst{7}
        \and G. Speranza\inst{6}
        \and I. García-Bernete\inst{8}
        \and A. Audibert\inst{1,2}
        \and A. Alonso-Herrero\inst{8}
        \and D. Rigopoulou\inst{9,10}
        \and A. Labiano\inst{3,11}
        \and J. A. Acosta-Pulido\inst{1,2}
        \and S. García-Burillo\inst{12}
       }

   \institute{Instituto de Astrof\' isica de Canarias, Calle V\'ia L\'actea, s/n, E-38205, La Laguna, Tenerife, Spain
    \and Departamento de Astrof\'isica, Universidad de La Laguna, E-38206, La Laguna, Tenerife, Spain
    \and Instituto de Radioastronom\' ia and Astrof\' isica (IRyA-UNAM), 3-72 (Xangari), 8701, Morelia, Mexico
    \and INAF – Osservatorio Astrofisico di Arcetri, largo E. Fermi 5, 50127 Firenze, Italy 
    \and Divisão de Astrofísica, INPE, Avenida dos Astronautas 1758, São José dos Campos 12227-010, SP, Brazil
    \and Instituto de Física Fundamental, CSIC, Calle Serrano 123, 28006 Madrid, Spain
    \and Department of Physics \& Astronomy, University of Sheffield, S3 7RH Sheffield, UK 
    \and Centro de Astrobiolog\' ia (CAB), CSIC-INTA, Camino Bajo del Castillo s/n, E-28692, Villanueva de la Cañada, Madrid, Spain
    \and Department of Physics, University of Oxford, Oxford OX1 3RH, UK
    \and School of Sciences, European University Cyprus, Diogenes street, Engomi, 1516 Nicosia, Cyprus 
    \and Telespazio UK for the European Space Agency, ESAC, Camino Bajo del Castillo s/n, 28692 Villanueva de la Cañada, Spain
    \and Observatorio Astronómico Nacional (OAN-IGN)-Observatorio de Madrid, Alfonso XII, 3, 28014, Madrid, Spain
    }

   \date{Received Feb 2, 2026; accepted Apr 13, 2026}
  \abstract
 {Merging luminous galaxies are ideal laboratories to study some of the most extreme astrophysical phenomena. The local ($z=0.1232$) obscured quasar J1356+1026 has two nuclei, North and South (J1356N and J1356S), but despite numerous efforts, J1356S had not yet been confirmed as an AGN. 
 Thanks to the superb sensitivity and spatial resolution of the MIRI/MRS instrument on board the JWST, we present new evidence suggesting that J1356S may indeed host an AGN with $\rm log~L_{\rm bol}=43.4\pm^{0.6}_{0.5}~erg~s^{-1}$. This is supported by the detection of strong coronal line emission at this location and by a spectral shape that differs from that of J1356N and those of the narrow-line region (NLR). 
 Aided by the spatially resolved information of MIRI/MRS and VLT/SINFONI, we also find that the high ionization gas, traced by the coronal lines [Ne~{\sc v}]$14.3~\mu$m and [Si~{\sc vi}]$1.963~\mu$m, has an extension of $\sim13-15.5$~kpc. This is likely a lower limit of the true extension, as suggested by the comparison with optical imaging from HST. The extended [Ne~{\sc v}] emission can be accounted for by photoionization from the quasar in J1356N in a relatively low density environment, ranging from $\rm n_e\leq 2000-3800~cm^{-3}$ in J1356N and $\rm n_e\leq 600-1200~cm^{-3}$ in J1356S and the NLR, as measured from the  [Ne~{\sc v}]$14.3~\mu$m and $24.3~\mu$m lines.}


   \maketitle
 \nolinenumbers
\section{Introduction}
The connection between merging galaxies and active galactic nuclei (AGN) in luminous hosts, oftentimes including an obscured phase, has long been regarded as having a fundamental role in shaping galaxy evolution \citep[e.g.][]{sanders96}. In such co-evolution scenario, the presence of dual AGN (two AGN separated by $\lesssim$10~kpc and sharing the same host galaxy)
during certain periods of time is a tangible possibility \citep{DeRosa19,Koss12}.
The confirmation of dual AGN is often done using X-ray data (e.g. \citealt{komossa03}) and/or emission line diagnostics obtained via high spectral and spatial resolution observations 
\citep[e.g.][]{u13,koss23,hermosa25}. An example of the latter are coronal lines \citep{rodriguez-ardila25}, whose ionization potentials (IPs) of $\gtrsim$100 eV make it unlikely for phenomena associated with star formation to ionize the atoms at such energies, although a contribution from shocks cannot be discarded \citep{contini01,Hernandez25}. 

\begin{figure*}[h!]
    \centering
    \includegraphics[width=0.27\linewidth]{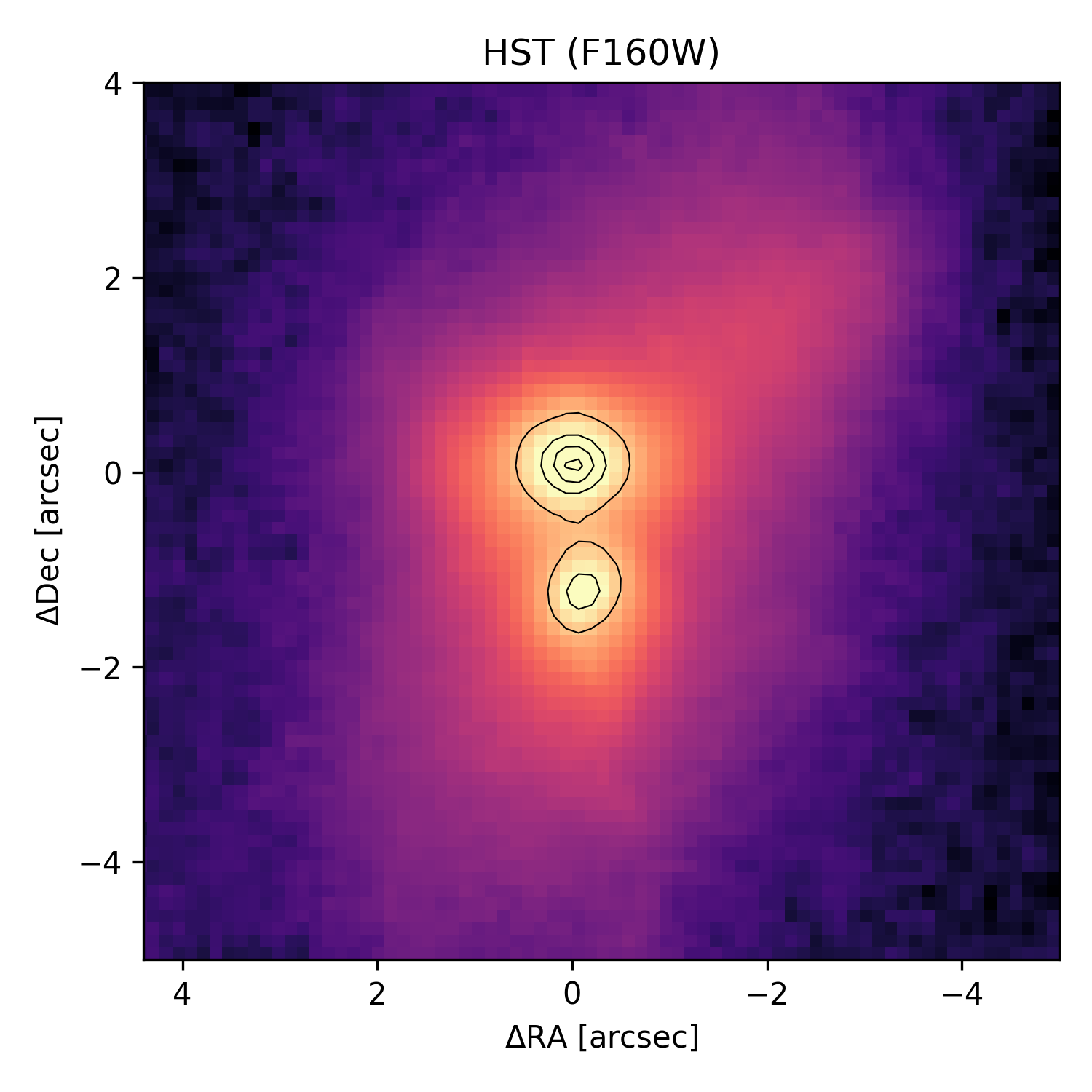}
    \includegraphics[width=0.7\linewidth]{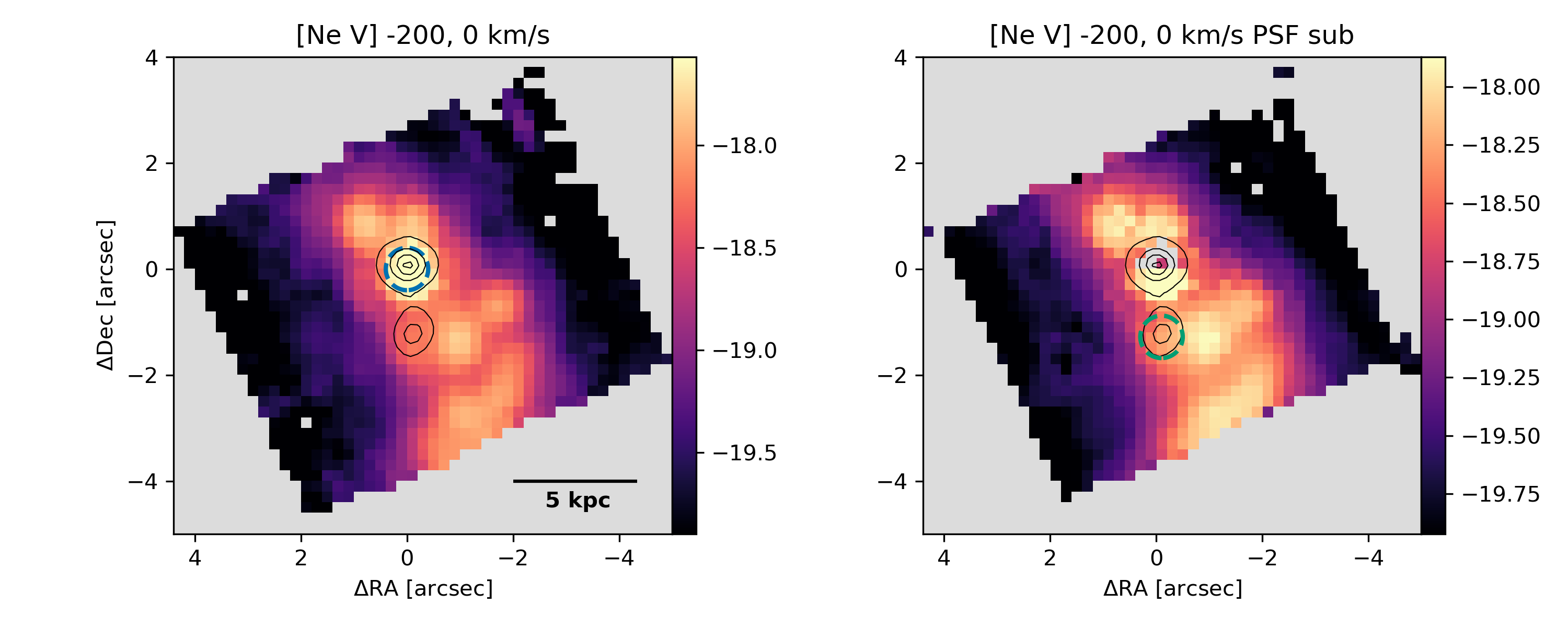}
    \caption{HST/WFC3 F160W image of J1356 in logarithmic scale (left) showing a similar region as the [Ne~{\sc v}]$14.3~\mu$m [-200, 0] \kms~velocity channel map obtained from the original (center) and PSF-subtracted (right) JWST/MIRI cubes. The zero velocity corresponds to the central wavelength of [Ne\,{\sc v}], redshifted to z=0.1232. The color bar is in units of erg~s$^{-1}$~cm$^{-2}$. The {[Ne\,{\sc v}]} emission extends to $\sim$6\arcsec~($\sim$13 kpc) along PA$\sim$30$^{\circ}$. The contours correspond to the HST F160W image, where J1356N and J1356S can be clearly identified. 
    The blue and green circles indicate the regions from which the spectra shown in Fig.~\ref{fig:spectra} were extracted.}
    \label{fig:coronal}
\end{figure*}

Type-2 quasars (QSO2s) are dust-obscured type-1 quasars frequently found in interacting and/or merging galaxies \citep{Pierce23}. These targets may represent a crucial, transition phase in the evolution of luminous galaxies, occurring between a gas-rich merger and a type-1 quasar phase \citep{Hopkins08}. A local example of this class of objects is SDSS J135646.10+102609.0, hereafter J1356. It is part of the Quasar Feedback (\href{https://research.iac.es/galeria/cristina.ramos.almeida/qsofeed/}{QSOFEED}) sample \citep{bessiere24}, hosted {in a galaxy merger} with log (L$_{\rm IR}$/L$_{\sun}$)=11.8 \citep{greene09,ramos-almeida22}. It shows a large-scale ionized outflow \citep{greene12} and two stellar nuclei separated by $\sim$1.31\arcsec~(2.9 kpc), as measured from Hubble Space Telescope (HST) F160W imaging \citep{comerford15}: the North nucleus, hereafter J1356N, where the QSO2 is located, and the South nucleus, J1356S, candidated to host another AGN.

Using {\em Chandra} data, \citet{comerford15} measured emission {at} >5$\sigma$ and 4.4$\sigma$ associated with the position of the two stellar bulges identified in the F160W image. However, they could not confirm J1356S as an AGN because of the surrounding diffuse soft X-ray emission, likely associated with the outflow. Deeper {\em Chandra} observations were used by \citet{foord20} with the same result. Both J1356N and J1356S are detected in cold and hot molecular gas \citep{sun14,ramos-almeida22,zanchettin25}, but so far there are no radio detections of J1356S in sub-arcsecond resolution data \citep{jarvis19,njeri25}. 
Beyond the complex nuclear region, J1356 has a small companion galaxy to the North ($\sim$57~kpc), a $\sim$20~kpc [O{\sc iii}] expanding bubble to the South \citep{greene12,speranza24}, and diffuse X-ray emission \citep{greene14,foord20}. 

The availability of JWST MIRI/MRS data of J1356, first published by \citet[hereafter {\color {blue}  RA25}]{ramos-almeida25}, allowed us to investigate both its possible dual AGN nature and extended coronal line emission, by means of several neon lines. 
We adopt a cosmology of H$_{0} = 70$~\kms Mpc$^{-1}$, $\Omega_m=0.3$, and $\Omega_{\Lambda}=0.7$. J1356 has a redshift of $z=0.1232$, corresponding to a luminosity distance of 575.8 Mpc and a spatial scale of 2.213~kpc/\arcsec. 

\begin{figure}
\includegraphics[trim={15 5 10 5}, clip=True,width=1\linewidth]{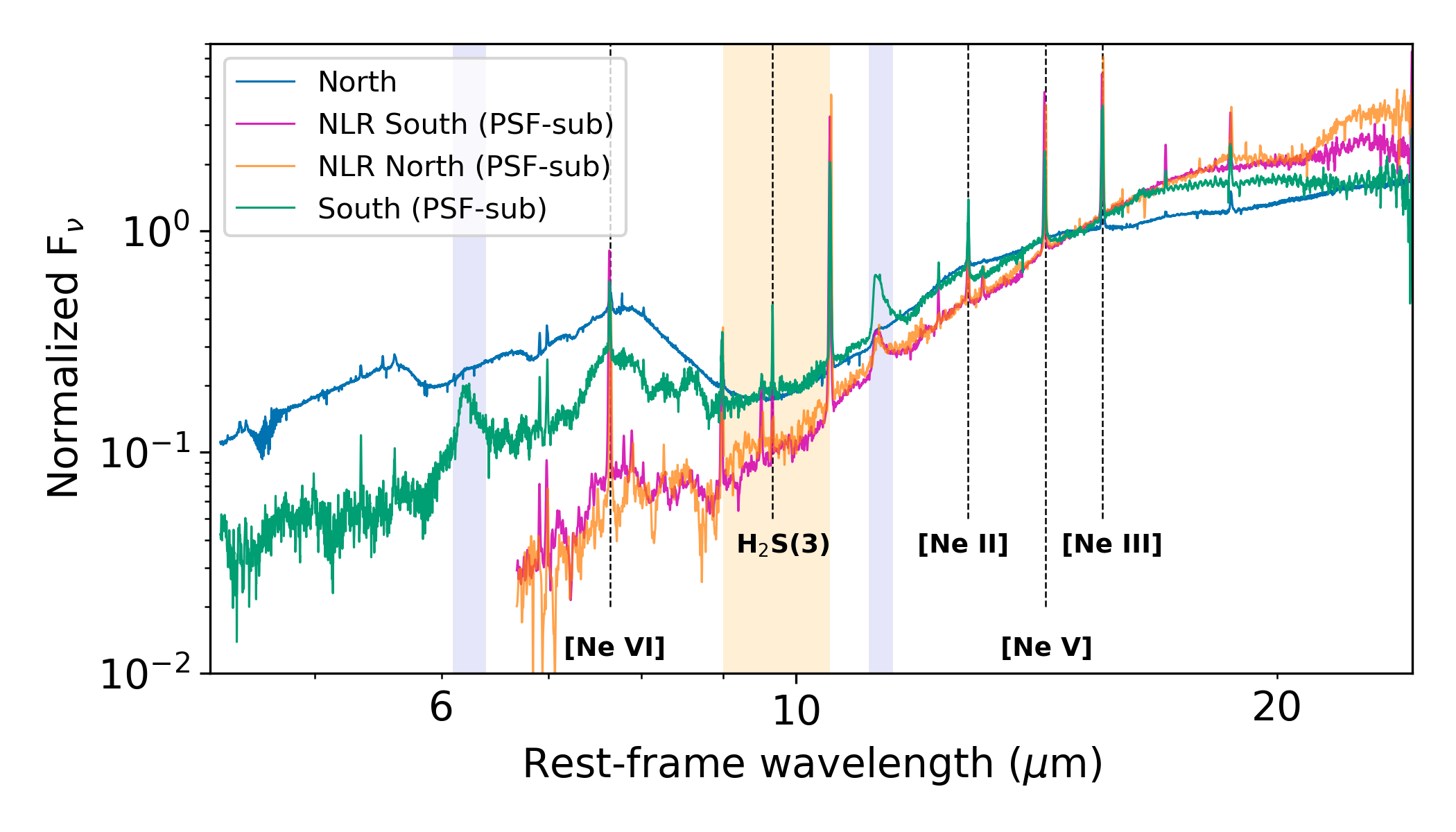}
    \caption{Spectra from J1356N (blue), J1356S (green), and two NLR locations (pink and orange) extracted in circular apertures of 0.4\arcsec~radius (shown in Fig.~\ref{fig:line_ratio}), normalized at 15 $\mu$m. The spectrum of J1356N was extracted from the original cubes, and the others from the PSF-subtracted ones. The spectra of the NLR could only be extracted from Ch2 to Ch4 due to the limited size of the Ch1 field-of-view (FOV).
   The emission lines used here are labeled. The purple and yellow shaded regions indicate the 6.2 and 11.3 $\mu$m PAHs and the 9.7 $\mu$m silicate absorption. For reference, the continuum fluxes at 15 $\mu$m of J1356N, J1356S, NLR S and N are 22.1, 1.31, 0.87, and  0.52 mJy.}
    \label{fig:spectra}
\end{figure}

\section{Observations and data reduction}
\label{sec:obs}
The data analyzed in this work are part of the JWST General Observer program 3655 (PI: Ramos Almeida; MAST \href{https://archive.stsci.edu/doi/resolve/resolve.html?doi=10.17909/8w9h-re72}{{doi:10.17909/8w9h-re72}}), previously published in {\color {blue} RA25}. 
The data of J1356 were taken on Jan 26th, 2025, using a 4- and 2-point dither sequence for the target's and background observations, respectively. We refer the reader to {\color {blue} RA25} for details on the observations and data reduction. 
Besides the standard data reduction, here we also used point spread function (PSF)-subtracted cubes. The PSF subtraction and associated data reduction were done as described in \citet{gonzalez-martin25}. This procedure is key for removing the contamination from the bright point source associated with J1356N (i.e., the QSO2), {allowing us to study} the underlying extended emission (see Fig.~\ref{fig:coronal}) and the spectrum of J1356S.

\section{The dual AGN nature of J1356}
\label{sec:dual}
Figure \ref{fig:coronal} shows the HST/WFC3 F160W contours overlaid on the [Ne{\sc v}]14.3$\mu$m channel maps, showing the position of J1356N and J1356S. We applied the routine \texttt{find\_peaks} from Astropy to determine the coordinates of the peak positions on the HST image, which are $1.28\arcsec$ (2.8~kpc) apart {in projection}. 
To extract the MIRI/MRS spectra of the two nuclei, we matched the peak of the local continuum around the [Ne{\sc v}] line to the position of J1356N in the HST/WFC3 F160W image. From that alignment, we determined the relative position of J1356S in the MIRI/MRS data (at $1.28\arcsec$ from J1356N). 
We then used the CAFE Region Extraction Tool Automaton \citep{cafe} to perform the 1D extractions in circular apertures of $0.4\arcsec$ radius, {matching} the angular resolution of Ch4.  
Fig. \ref{fig:coronal} shows the extraction apertures and Fig. \ref{fig:spectra} the corresponding spectra. The distinct nature of J1356N, whose mid-infrared spectrum was first reported in {\color {blue} RA25}, and J1356S, shown here for the first time, is evident from the different spectral shapes and relative line strengths (see Fig. \ref{fig:spectra} 
and Table \ref{tab:fluxes}). The intensities of the coronal lines, relative to lower ionization lines such as {[Ne{\sc ii}]12.8$\mu$m and [Ne{\sc iii}]15.5$\mu$m}, are higher in J1356S\footnote{The 
K-band spectra of J1356N and J1356S, shown in Fig. 1 in \citet{zanchettin25}, also show different slopes and [Si{\sc vi}] emission.}.
The 9.7 $\mu$m~silicate absorption feature is weaker in J1356S, whilst the polycyclic aromatic hydrocarbons (PAHs) are stronger than in J1356N. In addition, J1356S is clearly detected in H$_2$ (see bottom panel of Fig.~\ref{fig:h2_mom_map}). 

Despite the clear differences between J1356N and J1356S spectra, we do not see a point source in the MIRI/MRS continuum in the case of J1356S\footnote{The PSF-subtracted cubes show mid-infrared continuum emission from J1356S (see Fig. \ref{fig:cont_ch1}), which is also detected in the near-infrared (HST/WFC3 and VLT/SINFONI).}. 
This can be due to contrast (a relatively weak AGN embedded in a bright galaxy merger), but it is also possible that J1356S is a stellar nucleus whose mid-infrared spectrum shows projected narrow line region (NLR) emission from the QSO2 (J1356N). To test this possibility, we extracted spectra in two circular apertures of $0.4\arcsec$ radii centered in two locations dominated by the NLR (see Fig. \ref{fig:line_ratio}). The NLR emission was identified using the [Ne{\sc iii}]/[Ne{\sc ii}] map, following \citet{garcia-bernete24}. The slopes of the NLR spectra (pink and orange lines in Fig. \ref{fig:spectra}) are the same, but distinct from J1356N and J1356S. This, together with the location of J1356S outside the two projected ionization cones shown in Fig. \ref{fig:line_ratio}, suggests that it is not just part of the NLR of J1356N. 

\begin{figure}[h!]
    \centering
    \includegraphics[width=0.95\linewidth, trim={2cm 12.5cm 2.5cm 2cm}, clip=True]{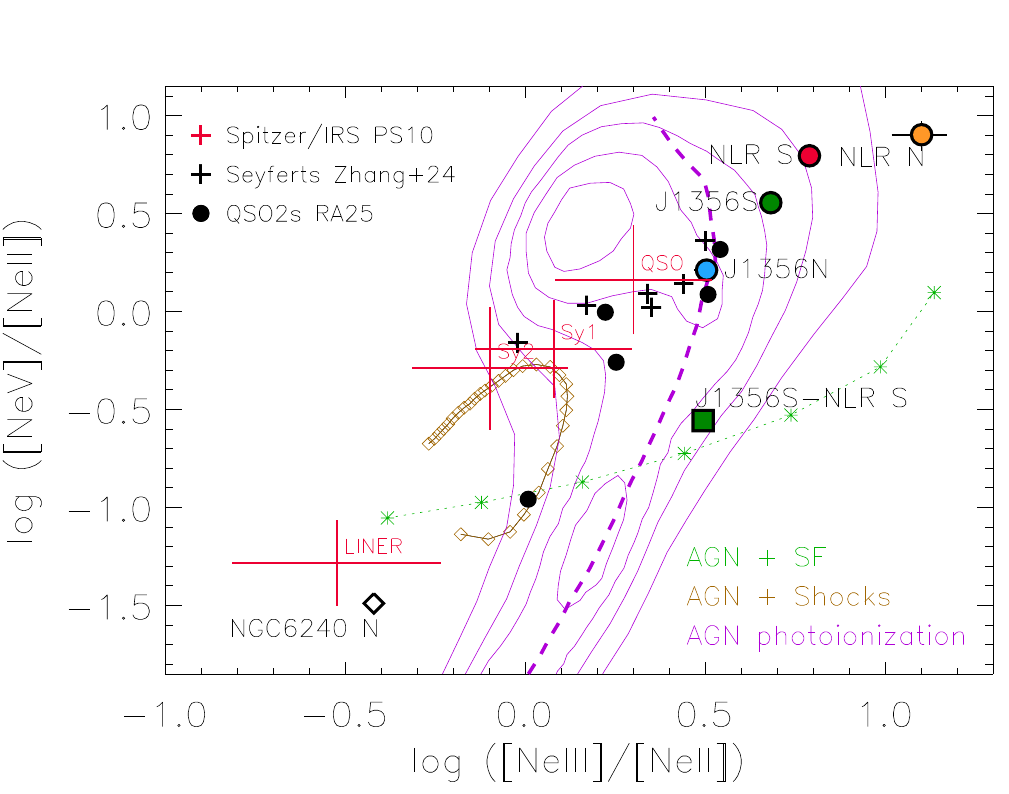}
    \caption{[Ne{\sc v}]/[Ne{\sc ii}] versus [Ne{\sc iii}]/[Ne{\sc ii}] diagram. The ratios corresponding to the spectra in Fig. \ref{fig:spectra} are shown as blue, green, orange, and pink circles. The NLR-subtracted ratios of J1356S are shown as a green square. 
    Black dots are the QSO2s in {\color {blue} RA25}, black crosses the Seyferts from \citet{zhang24}, and open diamond is the North nucleus of the dual AGN NGC\,6240 \citep{hermosa25}. Red lines indicate the median ratios of QSOs, Seyferts, and LINERs from {\color {blue} PS10}, measured from Spitzer/IRS spectra. Purple contours are the full grid of AGN photoionization models from \citet{feltre16}, 
    and green asterisks and brown diamonds are the AGN+SF and AGN+shocks models from \citet{feltre23}. See Appendix \ref{appendix} for details.}
    \label{fig:diagram}
\end{figure}

We measured the total flux of [Ne{\sc ii}], [Ne{\sc iii}], and [Ne{\sc v}] in J1356N and J1356S, and in the two NLR spectra shown in Fig. \ref{fig:spectra}  
see Table \ref{tab:fluxes}).  
[Ne{\sc v}]/[Ne{\sc ii}] is a diagnostic for nuclear activity, with AGN typically showing [Ne{\sc v}]/[Ne{\sc ii}]$>0.1$ \citep{inami13}, and [Ne{\sc iii}]/[Ne{\sc ii}] is sensitive to the hardness of the radiation field \citep{groves08}. Fig. \ref{fig:diagram} shows the [Ne{\sc v}]/[Ne{\sc ii}] vs. [Ne{\sc iii}]/[Ne{\sc ii}] diagram including the models from \citet{feltre16,feltre23}, computed for $\rm L_{bol}$=10$^{45}$~erg~s$^{-1}$ (see Appendix \ref{appendix} for details).  
The ratios of J1356N, J1356S, and the NLR South (NLR S) are consistent with these AGN photoionization models, as well as the five QSO2s studied in {\color {blue} RA25} and the Seyfert galaxies from \citet{zhang24}. 
There is a clear, positive trend followed by all data points, with the NLR spectra of J1356 showing the highest values of both ratios (see Table \ref{tab:fluxes}). 
We used {\it PyNeb} \citep{Luridiana15} with $\rm T_e$=10$^4$ and 2$\times$10$^4$~K, as in {\color {blue} RA25}, and the [Ne{\sc v}]14.3/24.3$\mu$m ratio to calculate the electron densities. This resulted in $\rm n_e\leq2000-3800~cm^{-3}$ for J1356N and $\rm n_e\leq$600-1200~cm$^{-3}$ for J1356S and NLR S (see Table \ref{tab:fluxes}). These values are consistent with the range of densities covered by the AGN photoionization models shown in Fig. \ref{fig:diagram}, of 10$^2$-10$^4$~cm$^{-3}$. The density decreases with distance from J1356N, with NLR N possibly having $\rm n_e\lesssim$100~cm$^{-3}$, according to its position in Fig. \ref{fig:diagram} and low [Ne{\sc v}]14.3/24.3$\mu$m ratio.

{Based on the JWST/MIRI observations of J1356S, it is possible that this stellar bulge hosts an AGN with lower luminosity than J1356N, but we cannot securely rule out that it is a star-forming galaxy whose spectrum shows projected emission from the NLR of J1356N. If we subtract the Ne line fluxes measured in the spectrum of NLR S from those of J1356S and plot the resulting ratios in Fig. \ref{fig:diagram} (green square), they are consistent with AGN photoionization only, but also with the AGN+SF model. However, both ratios are still higher than those reported for LINERs in \citet{Pereira10} (hereafter {\color {blue} PS10}). From the NLR-subracted [Ne{\sc v}]14.3$\mu$m flux of J1356S we estimate $\rm log~L_{\rm bol}\sim43.4\pm^{0.6}_{0.5}~erg~s^{-1}$} using Eq. 2 from \citet{spinoglio22}.  
For J1356N we measure log~L$_{\rm bol}$=45.4$\pm0.2$~erg~s$^{-1}$, 
consistent with the value of 45.3 erg~s$^{-1}$ measured from the extinction-corrected [O{\sc iii}] flux ({\color {blue} RA25}). 
{Using} the rest-frame intrinsic 2-7~keV luminosities reported by \citet{foord20} 
for J1356N and J1356S, and the correction of 20 from \citet{vasudevan07}, we obtain 
log~L$_{\rm bol}$={45.1}$\pm^{0.1}_{0.2}$ and {41.6}$\pm^{0.2}_{0.3}$ erg~s$^{-1}$, 
respectively. The LINERs with log L$_{\rm bol}$$\lesssim$43 erg~s$^{-1}$ in {\color {blue} PS10} show lower [Ne{\sc v}]/[Ne{\sc ii}] and [Ne{\sc iii}]/[Ne{\sc ii}] ratios than the NLR-subtracted J1356S values (see Fig. \ref{fig:diagram}), making L$_{\rm bol}$ estimated from [Ne{\sc v}] more likely to be representative of the possible AGN than that from the X-rays.

\section{The extended coronal line gas}
\label{sec:coronal}
 
A visual inspection of the MIRI/MRS data cubes revealed extended emission in the coronal lines, including [Ne{\sc v}]{ 14.3$\mu$m} and [Ne{\sc vi}]7.7$\mu$m (IPs = 97 and 126 eV). To investigate this further, we built continuum-subtracted velocity channel maps. 
Fig.~\ref{fig:coronal} shows the original and PSF-subtracted [-200, 0]~\kms~{channel map} of [Ne{\sc v}], which show an extended and clumpy gas distribution. These clumpy structures are also observed in the [Ne{\sc vi}] channel maps (see Fig. \ref{fig:cmap_nevi}), which cover a smaller FOV than the [Ne{\sc v}] maps. The maximum extension measured from the [Ne{\sc v}] maps, along a position angle (PA) of $\sim$30$^{\circ}$, is $\sim$6\arcsec~($\sim$13 kpc). 

The bright [Ne{\sc v}] emission at the southern edge of the FOV (detected at 6$\sigma$ and 4$\sigma$ in the original and PSF-subtracted cubes; see Figs. \ref{fig:coronal} and \ref{fig:cmap_nev}) suggests that it might extend even further.  
 A comparison with the HST/WFC3 F438W image of J1356 (see top panel of Fig. \ref{fig:nev_hst}) shows a clear correspondence between the [Ne{\sc v}] emission contours and the optical emission, mostly dominated by [O{\sc ii}]3726,3728\AA~in that filter (see the bottom panel of Fig. \ref{fig:nev_hst} for comparison, dominated by the near-infrared stellar continuum). Further supporting evidence for an even more extended [Ne{\sc v}] emission comes from the [Si{\sc vi}]1.963$\mu$m emission shown in the left panel of Fig. \ref{fig:h2_mom_map}, obtained from the VLT/SINFONI data first published by \citet{zanchettin25}. The [Si{\sc vi}] emission is detected at 3$\sigma$ at the southern edge of the SINFONI FOV, with an extension of up to $\sim$5\arcsec~to the South of J1356N and $\sim$7\arcsec~of total extension ($\sim$15.5 kpc). 
 Given that the IP of [Si{\sc vi}] (167~eV) is higher than that of [Ne{\sc v}], 
 it is reasonable to assume that the [Ne{\sc v}] has at least the same extension as the [Si{\sc vi}]-emitting gas. It is possible that the 
 coronal line emission might reach up to the 20~kpc ionized gas outflowing bubble first reported by \citet{greene12}.

In nearby Seyferts, coronal lines have been observed with extensions ranging from $\sim$100-200 pc \citep{riffel21} to up to $\sim$2-3~kpc \citep{rodriguez-ardila25}.
In local QSO2s, [Si{\sc vi}] emission extending up to $\sim$1 kpc has been measured {from} near-infrared spectra \citep{ramos-almeida17,ramos-almeida19,Speranza22}. So far, the maximum extent reported for coronal line emission is 23~kpc, measured for the [Fe{\sc vii}]3760\AA\ emitting-gas detected in MaNGA data of a galaxy merger at z=0.13 \citep{negus21}. However,  
[Fe{\sc vii}] is not detected in the nucleus, making it a good candidate for relic extended coronal emission. Thus, the projected size of the coronal emission of J1356, of 13-15.5 kpc, is among the largest ever observed.

To test whether shocks are required to explain the extended coronal emission in J1356 \citep[see][]{fonseca-faria23, kader26}, we calculated the [Ne{\sc v}]/[Ne{\sc ii}] and [Ne{\sc iii}]/[Ne{\sc ii}] ratios across the whole FOV of Ch3 and plotted them on Fig. \ref{fig:diagram_2d} together with the same models as in Fig. \ref{fig:diagram}. We find that we can reproduce the extended coronal emission with photoionization from an AGN with the luminosity of J1356N in a relatively low-density environment.

\section{Conclusions}
\label{sec:conclusions}
In this Letter, we report the finding of one of the most extended coronal line regions ever detected, traced by [Ne{\sc v}]14.3$\mu$m and [Si{\sc vi}]1.963$\mu$m, in the galaxy J1356, reaching a projected extent of 13-15.5~kpc. This extent is likely a lower limit of the true size of the coronal line region, set by the reduced FOV of MIRI/MRS. The large extension can be explained by photoionization from the quasar in J1356N and the relatively low density of the system: $\rm n_e\leq 2000-3800~cm^{-3}$ in J1356N, $\rm n_e\leq 600-1200~cm^{-3}$ in J1356S and NLR S, and possibly a lower density in NLR N.

We also report new evidence for the possible presence of an AGN with $\rm log~L_{\rm bol}=43.4\pm^{0.6}_{0.5}~erg~s^{-1}$ in J1356S, although we cannot rule out that it is a star-forming galaxy whose mid-infrared spectrum includes projected emission from the NLR of J1356N. Further comparison with low-luminosity AGN and stellar photoionization models, coupled with adaptive optics near-infrared IFU observations, might be required to confirm the presence of a dual AGN in this merger system.

\begin{acknowledgements}
This work is based on observations made with the NASA/ESA/CSA James Webb Space Telescope. The data were obtained from the Mikulski Archive for Space Telescopes at the Space Telescope Science Institute, which is operated by the Association of Universities for Research in Astronomy, Inc., under NASA contract NAS 5-03127 for JWST and from the European JWST archive (eJWST) operated by the ESAC Science Data Centre (ESDC) of the European Space Agency.  
MB acknowledges support from the Juan de La Cierva scholarship with reference JDC2023-052684-I, funded by MICIU/AEI/10.13039/501100011033 and FSE+. MB, CRA and AA thank the Agencia Estatal de Investigaci\'on of the Ministerio de Ciencia, Innovaci\'on y Universidades (MCIU/AEI) under the grant ``Tracking active galactic nuclei feedback from parsec to kiloparsec scales'', with reference PID2022$-$141105NB$-$I00 and the European Regional Development Fund (ERDF). OG-M acknowledges financial support from the SECIHTI project Ciencia de Frontera CF-2023-G100, UNAM/DGAPA project PAPIIT IN109123, and Estancias Sabáticas PASPA of UNAM/DGAPA. MVZ acknowledges the support from project "VLT-MOONS" CRAM 1.05.03.07. MC thanks the financial support from Coordenação de Aperfeiçoamento de Pessoal de Nível Superior – Brasil (CAPES) – Finance Code 001. IGB is supported by the Programa Atraccíon de Talento Investigador ``Cesar Nombela'' via grant 2023-T1/TEC-29030 funded by the Community of Madrid. MPS acknowledges support from grants RYC2021-033094-I, CNS2023-145506, and PID2023-146667NB-I00 funded by MCIN/AEI/10.13039/501100011033 and the European Union NextGenerationEU/PRTR. AA also acknowledges support from the European Union (WIDERA ExGal-Twin, GA 101158446).  {This work made use of \href{http://www.astropy.org}{Astropy}: a community-developed core Python package and an ecosystem of tools and resources for astronomy \citep{astropy:2018}. We thank the anonymous referee for constructive comments that helped improving this manuscript.}
\end{acknowledgements}

\bibliographystyle{bibtex/aa.bst}
\bibliography{refs.bib}

\begin{appendix}

\section{Supporting material}
\label{appendix}

This Appendix provides supporting evidence for the {possible} dual AGN nature and extended coronal line emission of J1356.\\

\noindent
{The models shown as purple lines in Fig. \ref{fig:diagram} correspond to the grid of photoionization models of AGN NLR from \citet{feltre16}, which were computed using the \texttt{CLOUDY} code (version c13.03; \citealt{Ferland13}) and the same parametrization of the metal and dust content in the ionized gas as in \citet{Gutkin16}. 
\citet{feltre16} chose an open geometry and a broken power law of spectral index $\alpha$ ranging from -2 to -1.2 to reproduce the emission from the AGN accretion disc, which is described in Eq. 5 there. They adopted a fixed AGN luminosity of $\rm 10^{45}~erg~s^{-1}~cm^{-2}$, an inner radius of the NLR of 300 pc, ionization parameter (U) in the range $\rm-4\leq\log U\leq-1$, fifteen values of the metallicity in the range $\rm0.0001\leq Z\leq0.07$, and dust-to-metal mass ratios in the range $0.1\leq\xi_d\leq0.5$. Finally, they considered hydrogen number densities (n$_{\rm H}$) ranging from 100 to 10~000~cm$^{\rm -3}$, which are consistent with the electron densities measured from the [Ne{\sc v}] lines except in the case of NLR N, where it is lower (see Table \ref{tab:fluxes}). The purple dashed line in Fig. \ref{fig:diagram} correspond to the photoionization model with n$_{\rm H}$=10$^3$~cm$^{\rm -3}$ (in good agreement with the electron density measured from the [Ne{\sc v}] line, of $\sim$1300 cm$^{\rm -3}$), Z=0.017, $\xi_d$=0.3, $\alpha$=0.7, and log($\langle U\rangle$) varying from -1.5 to -4.5 from top to bottom. Finally, the green asterisks and brown diamonds are the AGN+star formation (SF) and AGN+shocks models from \citet{feltre23}. They have 90\% contribution to the total H$\beta$ emission from star formation and shocks, respectively. In the case of the SF+AGN model, the ionization parameter increases from log($\langle U\rangle$)=-3.0 from left to right. While in the AGN+shocks model, the shock velocity goes from 200 to 1000 km~s$^{-1}$ counterclockwise from bottom to top. 

}

\begin{figure}[h!]
    \centering
    \includegraphics[width=0.9\linewidth]{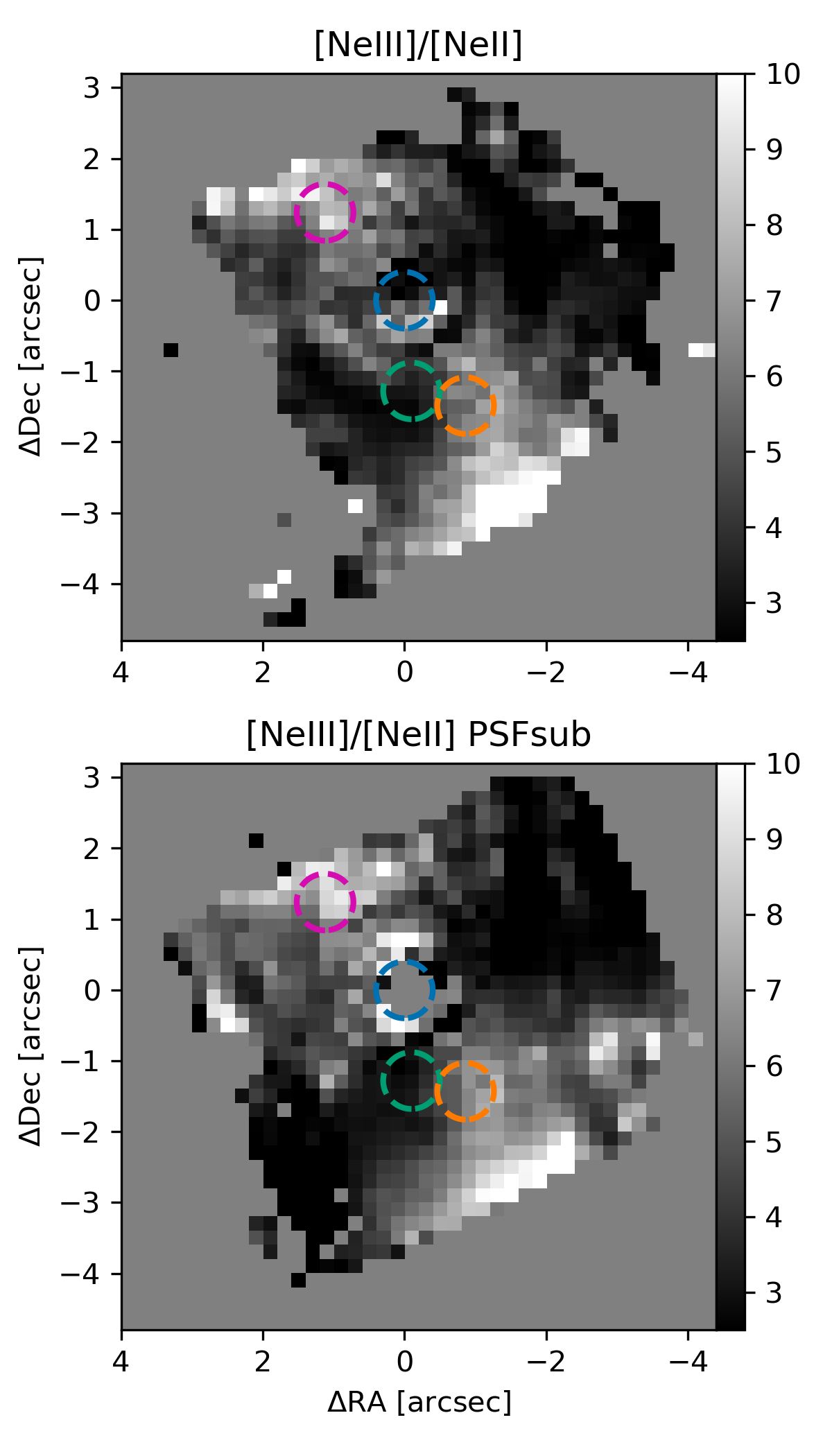}
    \caption{[Ne{\sc iii}]$15.5\mu$m/[Ne{\sc ii}]$12.8\mu$m emission line ratio measured from the original (top) and PSF-subtracted (bottom) cubes. The extractions corresponding to J1356N, J1356S, and the two locations within the ionization cones (NLR North and South) are indicated as blue, green, pink, and orange circles. }
    \label{fig:line_ratio}
\end{figure}

\begin{table*}[h!]
\caption{\label{tab:fluxes} Emission line fluxes of [Ne{\sc ii}], [Ne{\sc iii}], and [Ne{\sc v}] and corresponding ratios}
\centering
\begin{tabular}{lcccc}
\hline\hline
 & J1356N & J1356S & NLR N & NLR S \\
\hline
 [Ne{\sc ii}]   & 15.81$\pm$1.22 & 1.29$\pm$0.05 &0.18$\pm$0.03 & 0.72$\pm$0.02 \\
$[$Ne{\sc iii}] & 50.43$\pm$0.72 & 6.19$\pm$0.19 & 2.34$\pm$0.06 & 4.41$\pm$0.08\\
$[$Ne{\sc v}]14.3& 25.69$\pm$0.84 & 4.61$\pm$0.17 & 1.48$\pm$0.05 & 4.46$\pm$0.11\\
$[$Ne{\sc v}]24.3& 26.12$\pm$8.30 & 6.73$\pm$3.45 & 3.24$\pm$0.43 & 6.22$\pm$2.72\\
\hline
$[$Ne{\sc iii}]/[Ne{\sc ii}] & 3.19$\pm$0.25 & 4.81$\pm$0.23 & 12.61$\pm$2.19 & 6.16$\pm$0.20\\
$[$Ne{\sc v}]/[Ne{\sc ii}] & 1.62$\pm$0.14 & 3.59$\pm$0.19 & 7.97$\pm$1.40 & 6.22$\pm$0.23\\
$[$Ne{\sc v}]14.3/24.3 & 0.98$\pm$0.31 & 0.69$\pm$0.35 & 0.46$\pm$0.06 & 0.72$\pm$0.31\\
\hline
n$_{\rm e}$ (cm$^{\rm -3}$; $\rm T_e=10^4~K$) & $\leq$2008 & $\leq$606 & \dots & $\leq$606 \\
n$_{\rm e}$ (cm$^{\rm -3}$; $\rm T_e=2\times10^4~K$) & $\leq$3813 & $\leq$1221 & \dots & $\leq$1221 \\
\hline
\end{tabular}
\tablefoot{Fluxes are in units of $10^{-15}$\,erg\,s$^{-1}$\,cm$^{-2}$ and they were divided by a factor (1+z) because they were measured in the rest-frame spectra shown in Fig. \ref{fig:spectra}. We fitted the three emission lines with three Gaussian components plus a linear polynomial to describe the local continuum, using the \texttt{lmfit} package. {The last two rows correspond to upper limits on the electron densities measured from [Ne{\sc v}]14.3/24.3+$\Delta$[Ne{\sc v}]14.3/24.3 using \texttt{PyNeb} (v1.1.19) and considering electron temperatures of 10$^4$ and 2$\times$10$^4$~K. For reference, the median values of $[$Ne{\sc iii}]/[Ne{\sc ii}] and $[$Ne{\sc v}]/[Ne{\sc ii}] reported by \citet{Pereira10} for QSOs, Seyfert 1, Seyfert 2 galaxies, and LINERs, measured from Spitzer/IRS spectra, are shown in Fig. \ref{fig:diagram}. The median $[$Ne{\sc v}]14.3/24.3 values reported for the same groups are 1.00$\pm$0.30, 0.91$\pm$0.25, 1.00$\pm$0.30, and 0.77$\pm$0.18.}} 
\end{table*}

\begin{figure}[h!]
    \centering
    \includegraphics[width=0.7\linewidth]{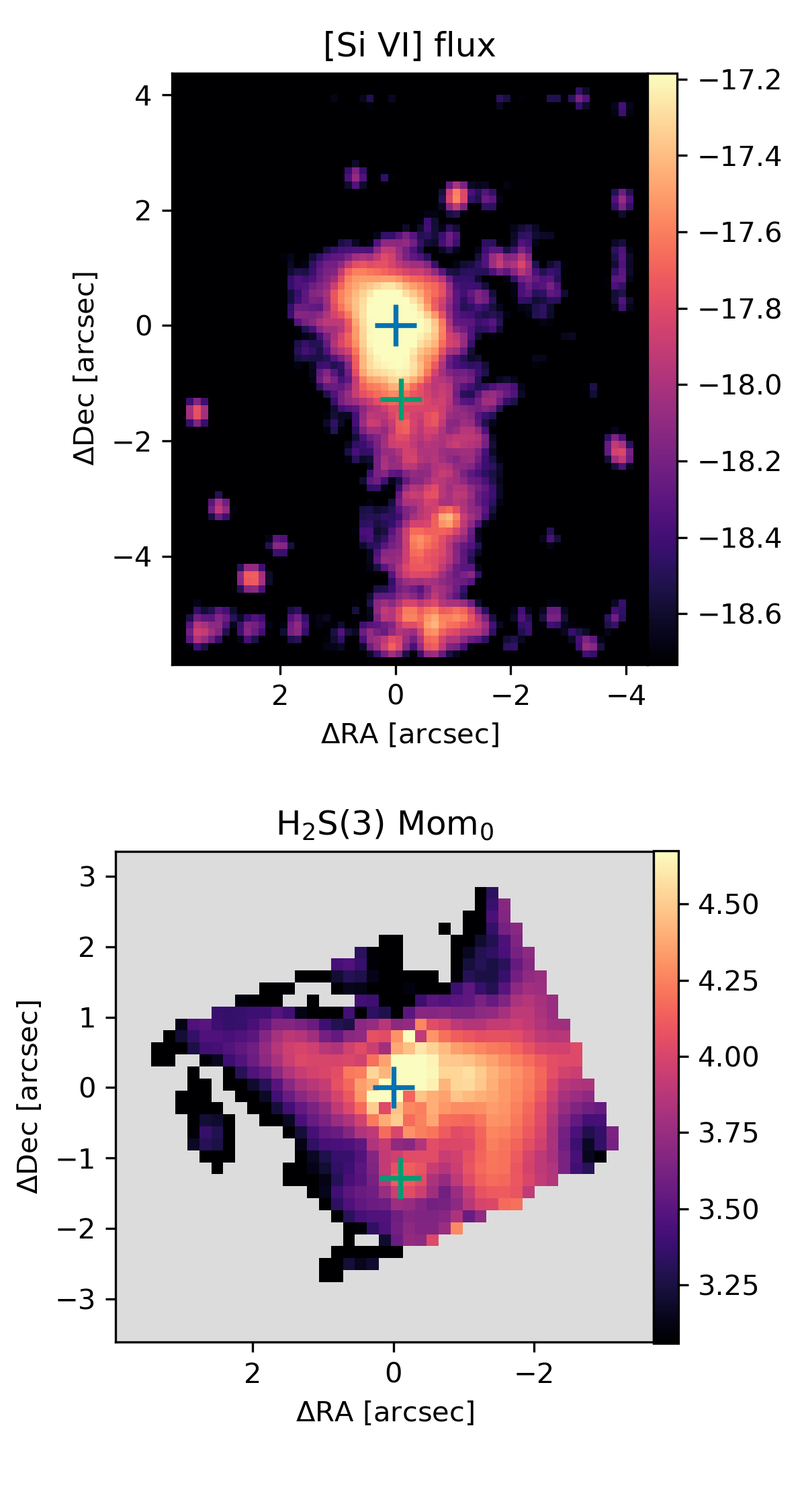}
    \caption{The top panel shows the {VLT/SINFONI} continuum-subtracted [Si{\sc vi}]1.963 $\mu$m flux map obtained from the fitting with one Gaussian component to the emission line. The color bar is in units of erg~s$^{-1}$~cm$^{-2}$. The bottom panel shows the {JWST/MIRI PSF- and}
    continuum-subtracted H$_2 0-0$S(3) moment 0 map, with the color bar in units of mJy~sr$^{-1}$\kms. The {crosses} indicate the position of the North and South nuclei. The South nucleus coincides with a clump of H$_2$. An in-depth analysis of the H$_2$ excitation and kinematics will be presented in {\color {blue} Zanchettin et al. (in prep.).}}
    \label{fig:h2_mom_map}
\end{figure}

\begin{figure}[h!]
    \centering
    \includegraphics[width=0.8\linewidth]{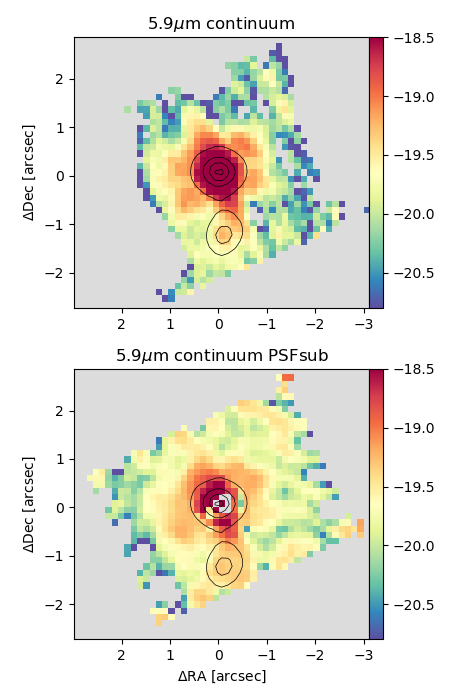}
    \caption{Local continuum centered at 5.9 $\mu$m in the original (top) and PSF-subtracted (bottom) cubes. The black contours correspond to the HST/WFC3 F160W image and indicate the location of J1356N and J1356S.}
    \label{fig:cont_ch1}
\end{figure}

\begin{figure*}[h!]
    \centering
    \includegraphics[width=0.9\linewidth]{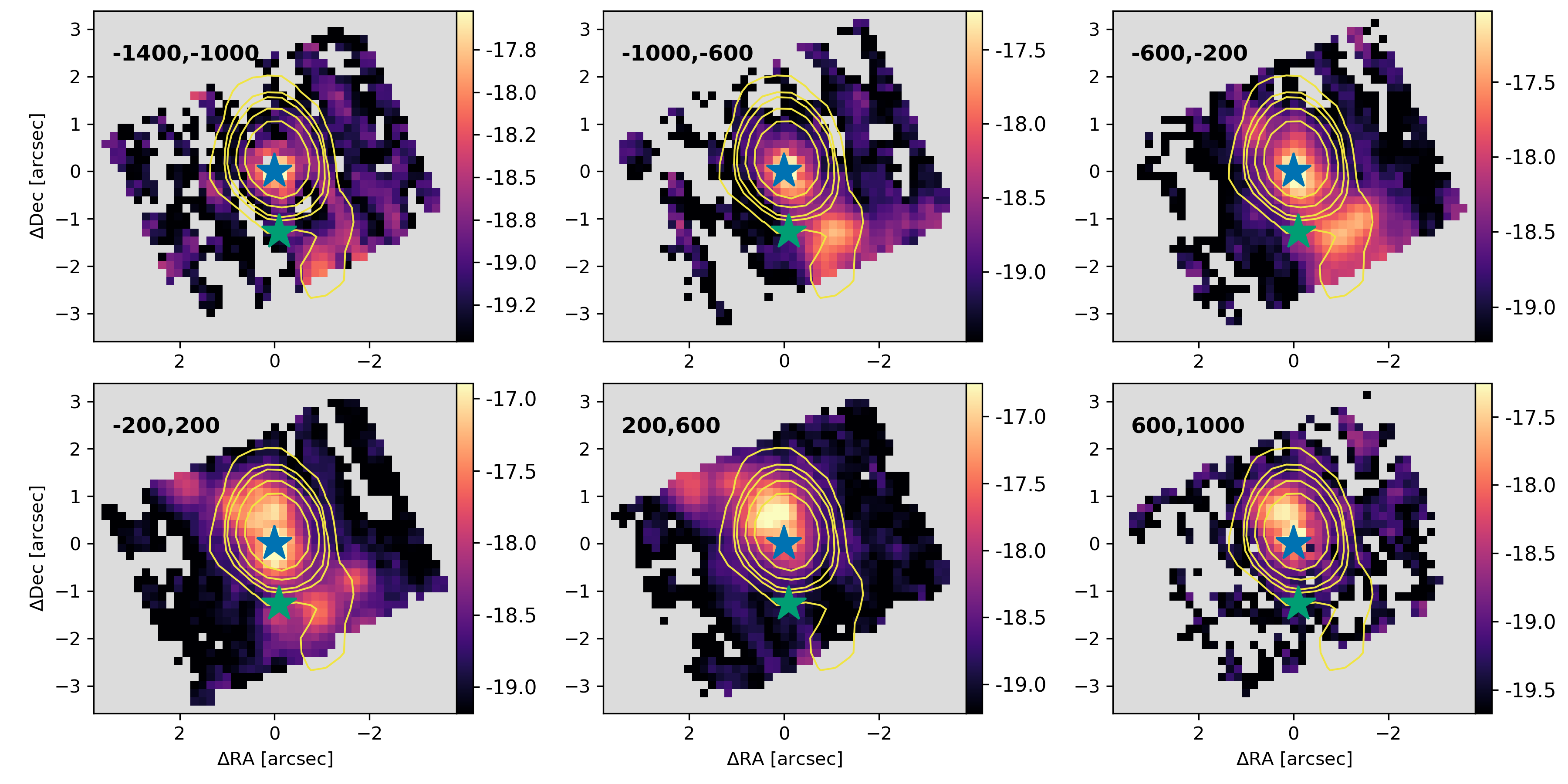}
    \caption{[Ne{\sc vi}]7.6$\mu$m velocity channel maps from -1400 to 1000~\kms, in increments of 400~\kms. The velocity slice used to produce each channel map is indicated on the top left corner of each panel. The zero velocity corresponds to the central wavelength of [Ne\,{\sc v}], redshifted to z=0.1232. The local continuum in each spaxel, modeled by a first degree polynomial, is subtracted from the line flux before building the velocity channels. The color bar is in logarithmic scale in units of erg~s$^{-1}$~cm$^{-2}$~pix$^{-1}$. The VLA 6~GHz contours from \citet{jarvis19} are overlaid, at levels of (3, 10, 15, 30, 60)$\times \sigma$. The stars mark the location of the N and S nuclei measured from the HST F160W image (see Fig. \ref{fig:coronal}).}
    \label{fig:cmap_nevi}
\end{figure*}

\begin{figure*}[h!]
    \centering
    \includegraphics[width=0.92\linewidth]{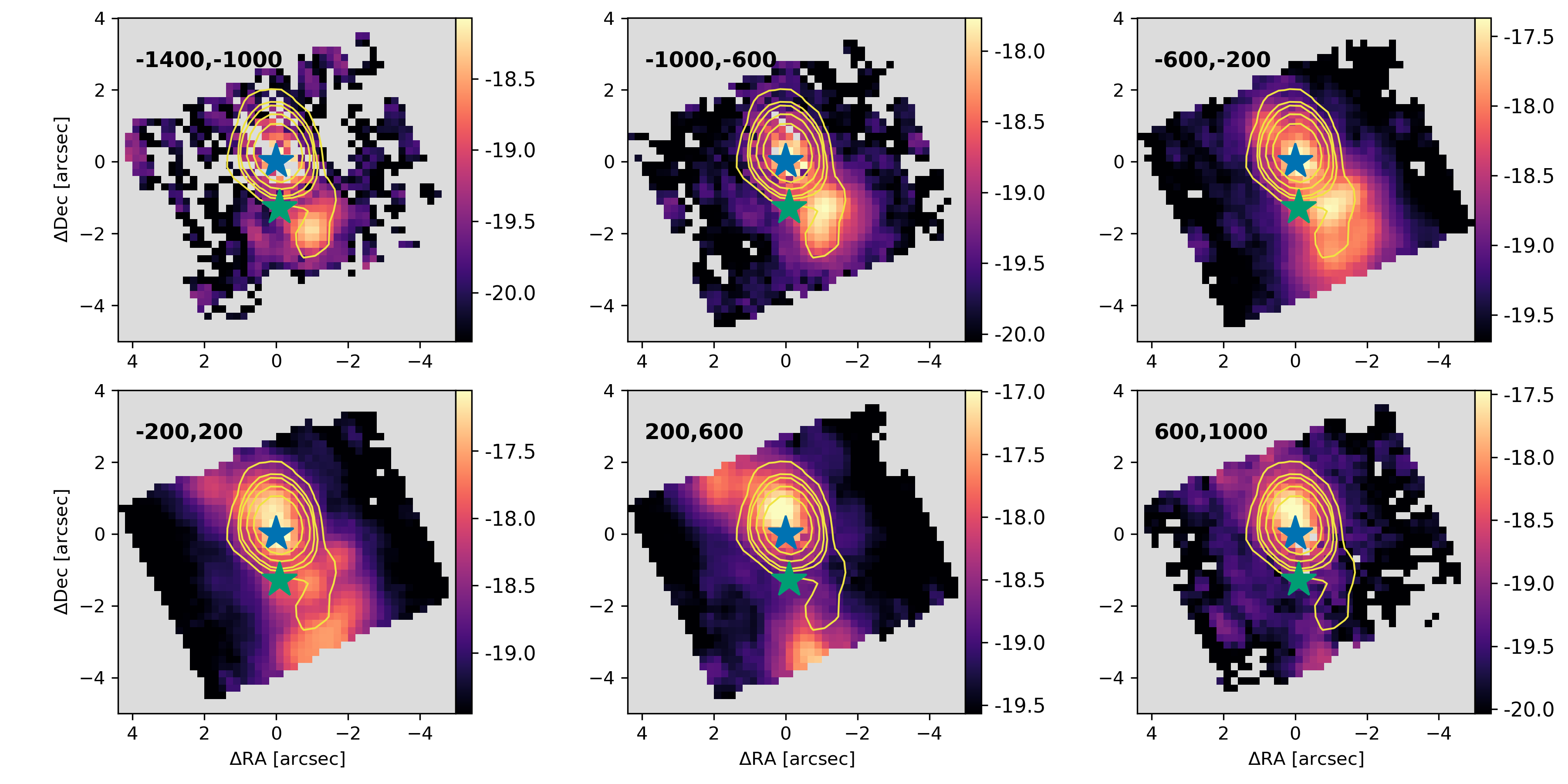}
    \caption{ Same as Fig.\,\ref{fig:cmap_nevi}, but for the [Ne{\sc v}]14.3$\mu$m emission.}
    \label{fig:cmap_nev}
\end{figure*}

\begin{figure}[h!]
    \centering
    \includegraphics[width=0.85\linewidth]{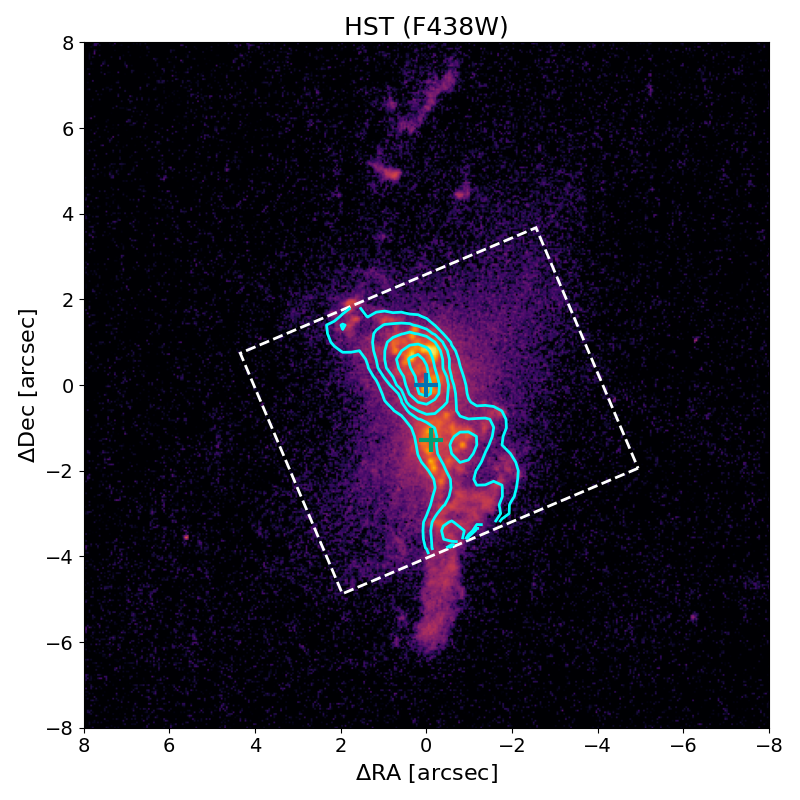}
    \includegraphics[width=0.85\linewidth]{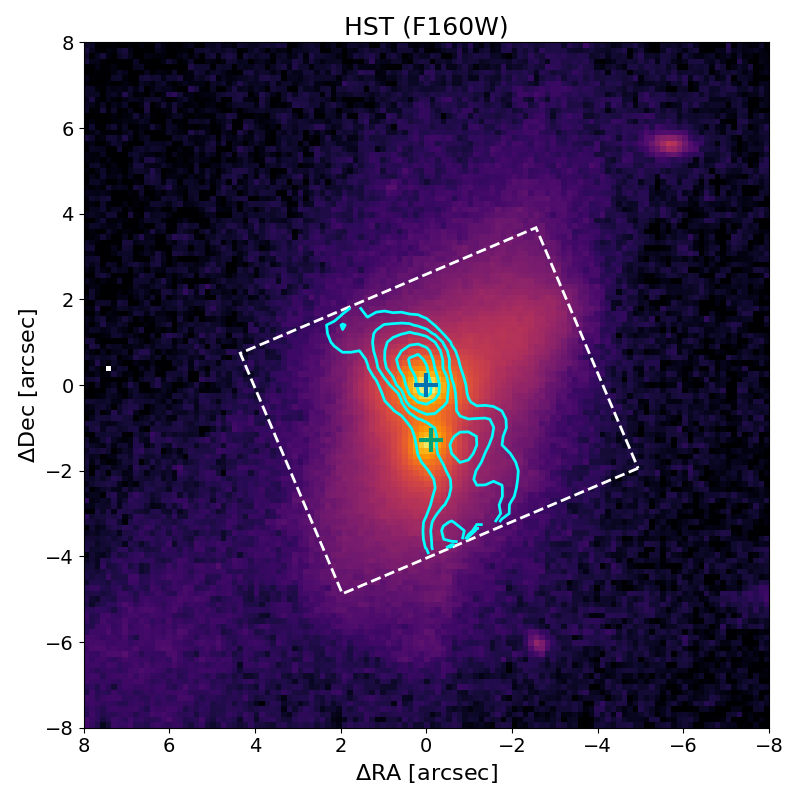}
    \caption{HST/WFC3 F438W and F160W images with the MIRI/MRS [Ne{\sc v}] flux contours overlaid in cyan. The dashed white box indicates the size of the MRS Ch3 FOV, and the crosses mark the position of the North and South nuclei.}
    \label{fig:nev_hst}
\end{figure}

\begin{figure}
    \centering
    \includegraphics[width=0.9\linewidth]{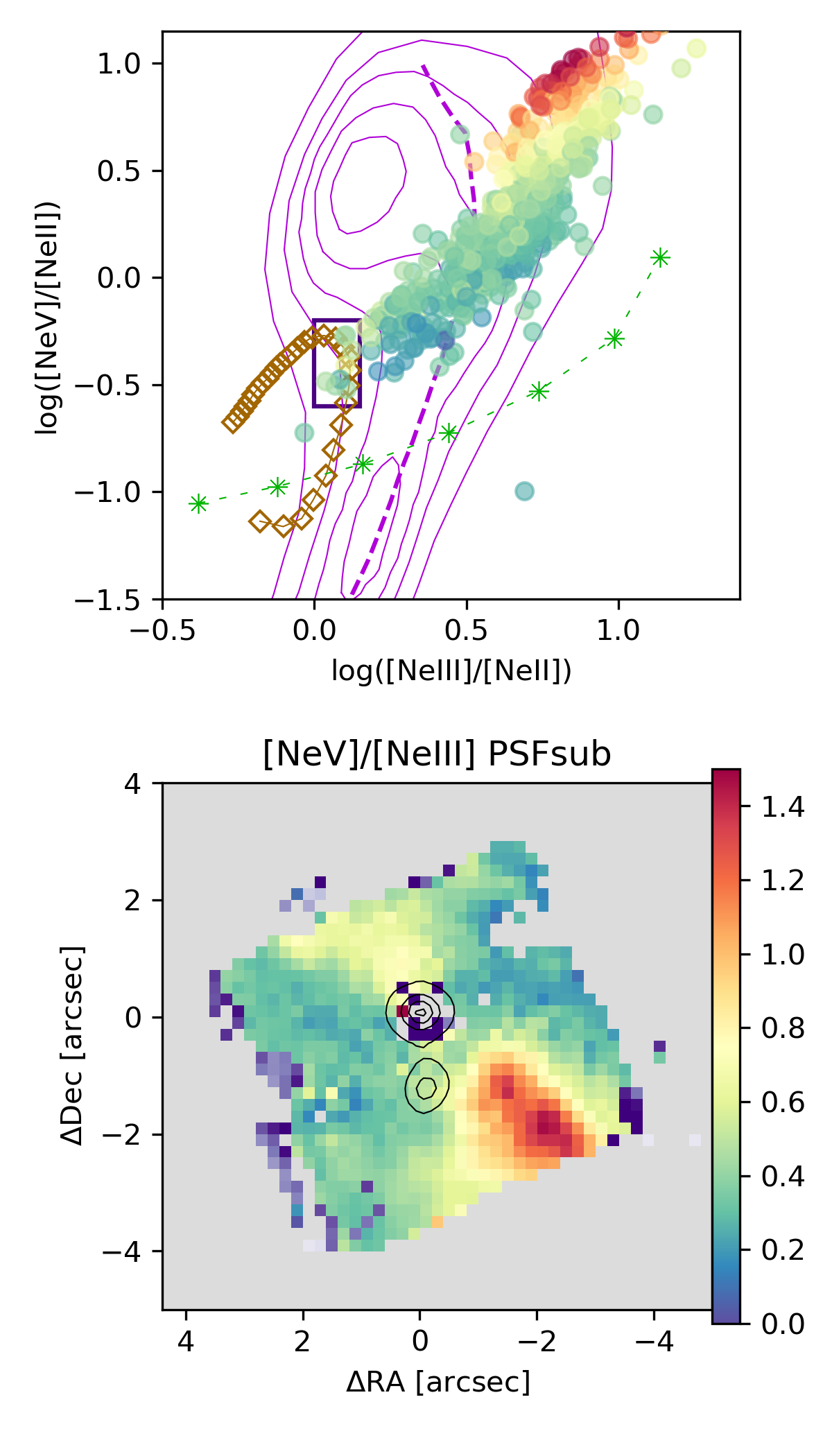}
    \caption{Same as Fig.~\ref{fig:diagram}, but for the spatially resolved measurements obtained from the PSF-subtracted cubes. The points are color-coded according to the [Ne{\sc v}]/[Ne{\sc iii}] line ratio shown in the bottom panel. The purple box ($0<\log($[Ne{\sc iii}]/[Ne{\sc ii}]$)<0.15$ and $-0.6<\log($[Ne{\sc v}]/[Ne{\sc ii}]$)<-0.2$) indicates the points closer to the AGN+shocks models (golden diamonds). Their spatial location is indicated by the purple areas on the bottom panel.}
    \label{fig:diagram_2d}
\end{figure}

\FloatBarrier 
\twocolumn

\end{appendix}
\end{document}